# Contrastive study on the single-file pedestrian movement of the elderly and other age groups

Xiangxia Ren[1], Jun Zhang[1*], Weiguo Song[1],

[1]State Key Laboratory of Fire Science, University of Science and Technology of China, Hefei 230027, China

**Abstract**:
The worldwide population is aging and countries are facing ongoing challenges in improving the safety of elderly pedestrians. In this work, single-file movement of the elderly are experimentally compared with that of different age groups. The findings indicates that the age is not the only factor influencing the pedestrian dynamics but the heterogeneity of the crowd composition and the familiarity among neighboring pedestrians also have significant effects. The existence of three regimes in the relationship between headway and speed is confirmed. In the strong constrained regime, the slope of the relationship between headway and speed of the elderly is bigger than that of the young, which means that the elders are more sensitive to the spatial headway than the young when adapting the speeds. However, the difference of the slopes in the weakly constrained regime is small, which indicates a weak dependency between age and the adaption time. The elderly need longer headway during the transformation of the motion state. Besides, the 'active cease' behavior of pedestrians, which is explained with the least effort principle, is observed in the experiment. The findings offer empirical data of the elderly under high densities and can be useful for the improvement of the pedestrian modelling and the construction of elderly friendly facilities.
**Keywords:** single-file movement, elderly crowd, headway, stop and go, active cease

1. Introduction

   Studies on the pedestrian movement, which can help to realize safer transportation environments and effective evacuation in emergencies, arouse growing concerns. Compared with young people, elders with mobility impairments and disease increment are in face with higher risks in their daily life especially in the transportation system. It is of great significance to study the movement properties of elderly pedestrians for the aging of population is a social phenomenon around the world.

   Experiments under well controlled but various conditions were carried out previously [1-10] to study the characteristics of pedestrian movement by considering different influence factors such as age, gender, building structure, movement motivation, data collection method and so on. However, pedestrian crowd is a complex system affected by several factors including experiment setup, facility geometry, and pedestrian movement motivation and so on. Experiments [11-16] on single-file pedestrian movement can reduce the influences of experiment conditions and the lateral interactions among pedestrians and focus on the influence of certain factors on the characteristics of pedestrian movement. The researches on pedestrian single-file movement mainly focus on time-space diagram, fundamental diagram and the relationship of headway and speed, etc. Zeng et al. [17] revealed the relationship between step length and step frequency under different headways. Cao et al. [18] studied the influence of the limited visibility on the single-file movement and found that the speed distribution in different conditions conforms to Gaussian distribution. Gulhare et al. [19] studied the differences of the single-file motion in the field observations and controlled experiment. They noted that the boundary might have an influence on the results.

The discrepancies of the basic fundamental diagrams of pedestrian movement [20,21] were investigated in different aspects including the shape of the walking path [14], age composition [11, 13, 15, [22], as well as cultural difference [23] with series of experiments. Seyfried et al. [11] measured the fundamental diagrams for the densities up to 2 $m^{-1}$ and found that two differing speed phases. In France, Jelic et al. [14,24] conducted experiment inside a ring and the density reached 3 $m^{-1}$. Beside the free flow regime and congested regime, they observed a third regime named weakly constrained regime in the relation of headway and speed. However, from the experiment of pedestrians with different ages in [15] in China three regimes in mixed group of old and young people but only two regimes in young students group were observed. This indicates that the heterogeneity of the crowd may lead to different characteristics of pedestrian flow. Unfortunately, the quantitative relation of headway and speed for the old group in the strong constrained regime were not presented due to the lack of data for old people movement especially at the high densities. It is obvious that most of the experiments focused on young pedestrians and the studies on elderly movement is still not enough.

Besides, the stop-and-go wave observed under high densities situations reproduced in several laboratory experiments [15,25,26] and simulation models [27]. Portz et al. [27-29] analyzed the stop-and-go waves by experiment and modeling qualitatively. An adaptive velocity model with reaction time was also proposed in [28], which is able to reproduce the stop-and-go waves qualitatively as observed in the experiments. Fang et al. [29] verified the similar behavior between pedestrian and vehicle, proposed the slow reaction (SR) model to describe the pedestrian's delayed reaction and reproduced the phenomena of uneven distribution in single-file movement. However, the stop-and-go waves of the elderly population are less studied and further experimental data are still required to verify the abovementioned models.

Based on these considerations, we performed laboratory experiments to investigate the movement of elderly pedestrian under different densities with the emphasis of high density. The aim of our study is to compare the characteristics of elderly group with that of the groups with different age compositions to understand the influence of age on the pedestrian movement. The rest of this paper is structured as follows: The setup of the experiment is described in section 2. The trajectories extracted from the experiment videos are displayed and the effects of different boundaries are discussed in section 3. Then we introduce the measurement methods and analyze the results in section 4. Finally, section 5 summarizes the paper and makes a conclusion.

**2. Setup of experiment**

The experiments were carried out in March 2018 in Hefei, China. Totally 73 volunteers without physical problems for normal movement were recruited from a senior center in Hefei. As shown in FIG.1, the mean age of them is 69.7±7 years old ranging from 52 to 81 years old and the ratio of the male to the female is about 1:2.5 (21 males and 52 females). The heights of the participants range from 150 cm to 175 cm with an average of 163 cm.

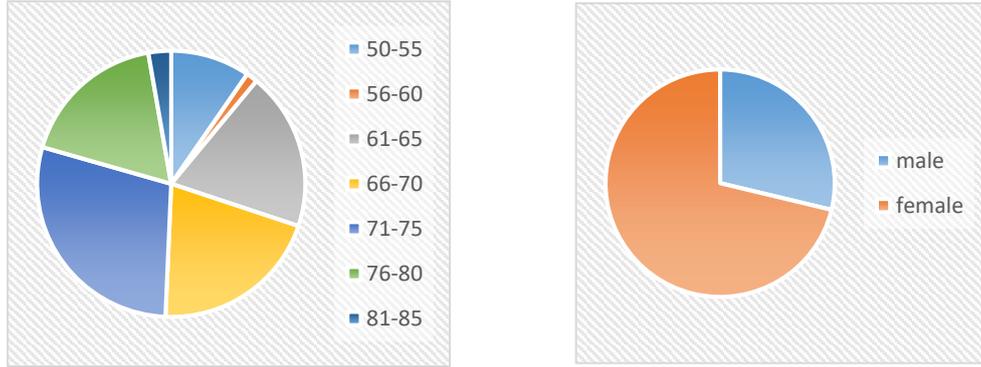

FIG.1. The distribution of age (left) and gender (right) of the pedestrians in the experiments.

FIG.2 shows the illustration and a screenshot of the experimental scenario, which is composed of two 5 m long straight corridors and two semicircle corridors with the inner radius 2.1 m and outer radius 2.9 m. The width of corridors was set as 0.8 m to avoid overtaking during movement. These experimental setups were the same as the experiment setup in [15] to make easy comparison. The circumference of the central line of the corridor, which was supposed as the walking route of the test persons, is about 25.7 m. The area 1 is a semi-closed boundary formed by setting up 1.8 m high boards at outer side of the corridor, while the area 3 is a closed boundary constructed with the boards from both sides to prevent pedestrians from crossing the border. For the area 2 and 4, the rough lines marked on the land surface were used to bind the path of pedestrians. At the beginning, the participants distributed along the 0.8 m wide oval circuit uniformly. Then they went ahead at their normal pace when we gave the order for the experiment to begin. As shown in TABLE I, six runs with different numbers of the participants in the corridor were performed to form different global density. Considering the limited length and patience, we only realize one time for each run.

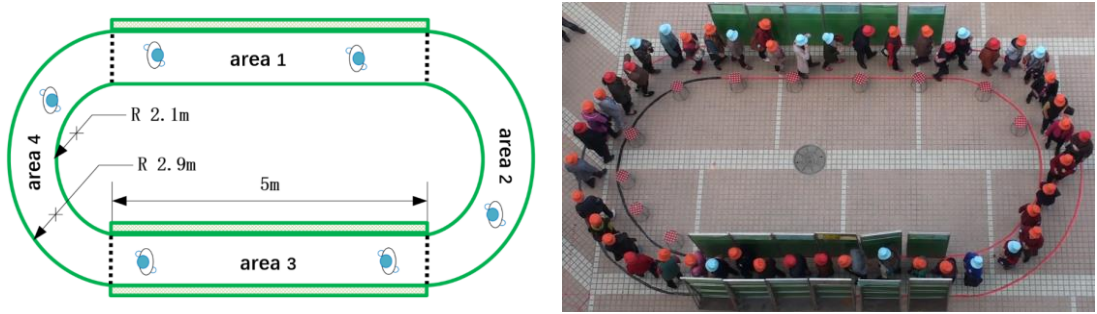

FIG.2. The sketch (left) and a screenshot (right) of the scenario, we changed the number of pedestrian to form different densities in the single-file.

TABLE I. Details of the runs of our experiment.

| Index | Name | Number of elders(N) | Duration time(s) | Global density $\rho_g$ [m$^{-1}$] |
|---|---|---|---|---|
| 01 | Elders-08 | 8 | 123 | 0.31 |
| 02 | Elders-20 | 20 | 95 | 0.78 |
| 03 | Elders-30 | 30 | 120 | 1.17 |
| 04 | Elders-39 | 39 | 95 | 1.52 |
| 05 | Elders-48 | 48 | 160 | 1.87 |
| 06 | Elders-56 | 56 | 148 | 2.18 |

Two digital cameras mounted on the roof of a building about 10 m high were used to record these experiments. Each participant wore a red or blue hat for easy recognition from video recordings (see

FIG.2). The software *PeTrack* [30] was used to extract the trajectories automatically and the average height of 163 cm was used for data transformation from pixel coordinates to physical coordinates. The error of the extracted position given by the software is within 0.1m [31], which is caused by insufficient calibration (0.05m in maximum), colored caps marker (0.02m in maximum) and the pedestrian height selection (0.02m in maximum).

### 3. Trajectories and the effect of the boundary

By means of digital image processing, we obtained the pedestrian trajectories under different scenes. In order to study the one-dimensional characteristics of single-file pedestrian movement further, the method mentioned in [25] is used to convert the original oval scenarios to straight ones of approximated 25.7 m long and 0.8 m wide. In the new coordinate system, y=0 corresponds to the central line of the corridor, and y<0 (resp. y>0) corresponds to the area which was inside (resp. outside) this oval central line before the unfolding. We did not filter out the steps before the study any more.

The converted trajectories with instantaneous speeds which are indicated by the color coded are displayed in FIG.3. In area 1, pedestrians are inclined to the side without a vertical boundary and keep a distance with the wall, while the pedestrian trajectories in area 3 are more concentrated and less fluctuant due to the constraint of the board boundaries on both sides. In area 2 and 4, the trajectories show relatively large fluctuations due to the lack of constraint effect of vertical boundaries. Some pedestrians even crossed the boundary and the occurring frequency increased with the increasing density in the corridor, which is caused by the overlapping of pedestrians under high density. Interestingly, the trajectories show a certain radian in the semicircle paths. FIG.4 (a) shows the graphic analysis of trajectory on the curve. Pedestrians are closer to the inner side of the curved corridor at the beginning of turning *B* while move outward gradually. After arriving around the vertex *H*, pedestrians no longer continue to move outward but move inward gradually. One of the reasons may be that pedestrians tend to increase the radius of their path to reduce the effect of inertia on the turning of the body. When approaching the vertex H of the semicircle, pedestrians made another path planning and move inward gradually to enter into the straight corridor by the shortest path. This may be a strategy for pedestrians moving on a curve to adapt to the constantly changing direction, which still needs further experimental data to proof.

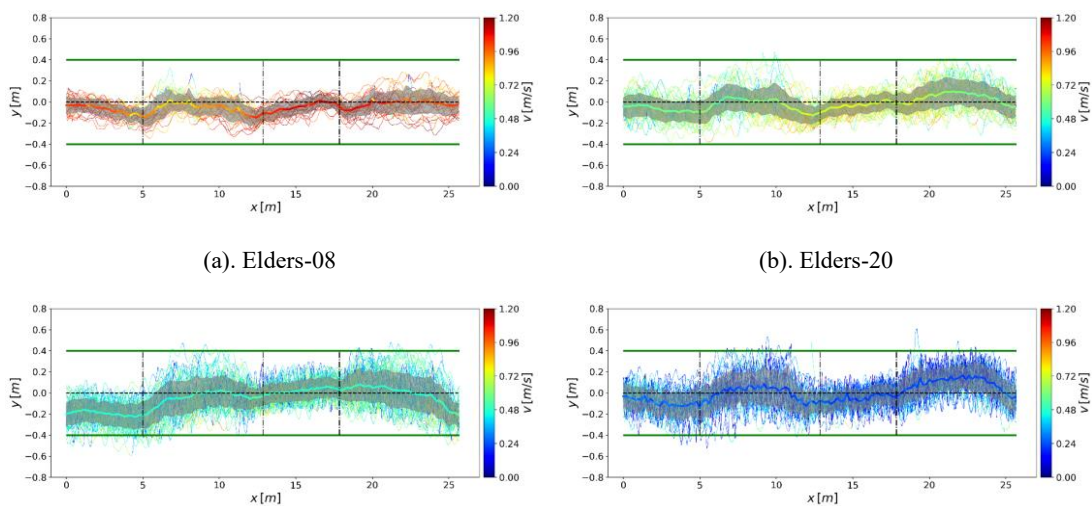

(a). Elders-08    (b). Elders-20

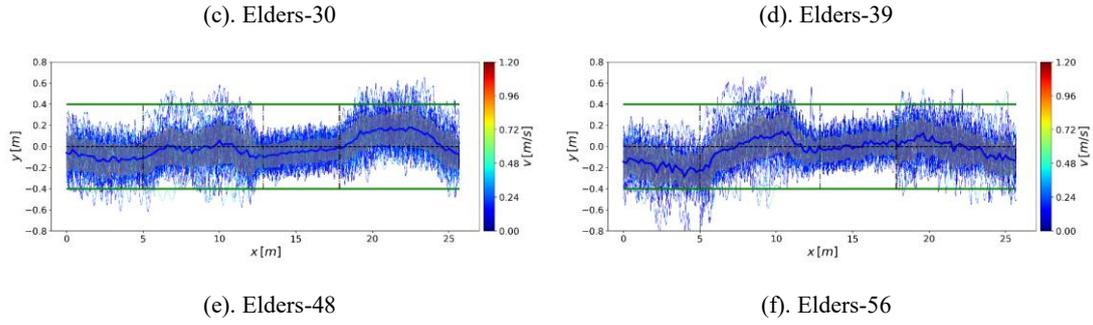

(c). Elders-30      (d). Elders-39

(e). Elders-48      (f). Elders-56

FIG.3. The adjusted trajectories with instantaneous speed which is indicated by color in the new coordinate system. The horizontal dotted line represents the center line of the corridor and the three vertical ones are used to distinguish the different regions. From left to right are area 1, area 2, area 3 and area 4. The thicker lines are obtained by averaging y values at 0.1 m intervals in the x direction and the shade represents the standard deviation.

At the first sight, there are obvious differences of the speeds in different areas of the scenarios Elders-08 and Elders-20 whose global densities are relatively low. Quantitatively, we further calculate the mean speeds with the standard deviation of different areas in each run as displayed in TABLE II and FIG.4 (b). The T-test is applied on the SPSS platform to test the significance of the difference and $p<0.05$ represents that there is a statistically significant difference between the samples. The detailed results of the P values are listed in TABLE III and $p<0.05$ for all of the T-test. It can be concluded that the shape of the path and the boundaries do affect the speed of pedestrians but mainly at low-density situations. The speeds at the curve parts are relatively lower than that in the straight part. Interestingly it is higher in the straight area with boundaries at both sides. The reasons may be as follows: as interpreted in [32], the wall of structure has a repellent effect on pedestrians. When pedestrians walk in the narrow corridor area 3, the repulsive force from vertical walls in both sides makes them uncomfortable and thus accelerate to leave there quickly, which is also confirmed by the fact that trajectories of pedestrians in area 1 are inclined to the side without wall. While when pedestrians enter the curve parts with relatively higher speeds, they need to decelerate to turn their bodies and keep balance during the continuous change of the movement direction. Otherwise, if they enter the curved section with lower speed under high densities, it is easy to keep balance and deceleration is not so necessary any more.

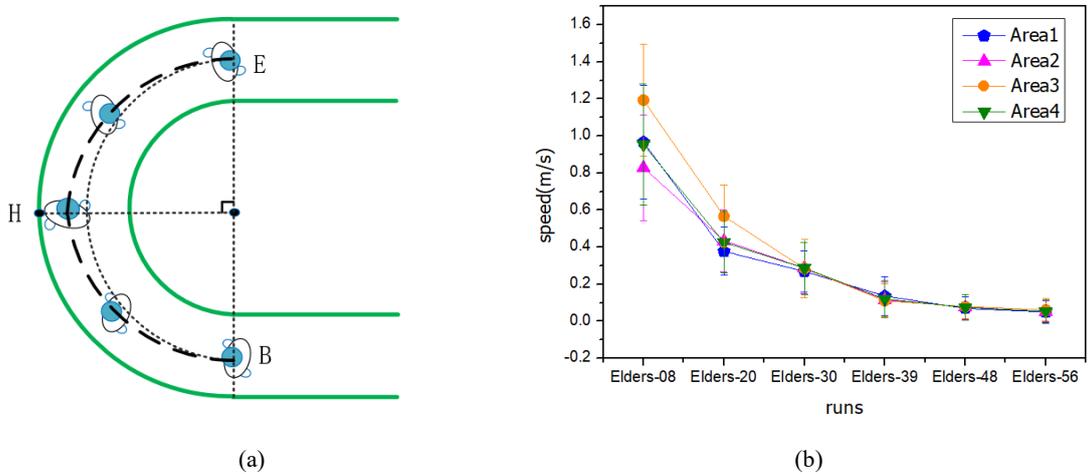

(a)      (b)

FIG.4 The graphic analysis of trajectory of pedestrians on the curve (a) and the mean speeds with the standard deviation of different areas in each run (b).

TABLE II. The mean speeds with the standard deviation of different measurement areas in each run.

| Runs | Elders-08 | Elders-20 | Elders-30 | Elders-39 | Elders-48 | Elders-56 |
|---|---|---|---|---|---|---|
| Area1 | 0.97±0.30 | 0.38±0.13 | 0.27±0.11 | 0.14±0.11 | 0.07±0.06 | 0.05±0.06 |
| Area2 | 0.83±0.29 | 0.43±0.17 | 0.28±0.14 | 0.12±0.09 | 0.08±0.07 | 0.05±0.06 |
| Area3 | 1.19±0.30 | 0.57±0.17 | 0.28±0.16 | 0.11±0.09 | 0.08±0.06 | 0.06±0.06 |
| Area4 | 0.95±0.33 | 0.43±0.17 | 0.29±0.14 | 0.12±0.10 | 0.08±0.07 | 0.05±0.06 |

TABLE III. The detailed results of P values in the T-test

| Index | Area3 and 1 | Area3 and 4 | Area1 and 4 |
|---|---|---|---|
| Elders-08 | 0.000 | 0.000 | 0.014 |
| Elders-20 | 0.000 | 0.016 | 0.000 |

## 4. Analysis and results

### 4.1 Fundamental diagrams

#### 4.1.1 Measurement methods

In this study, we analyze the variables like density and speed from both micro and macro aspects. At the micro level, the individual density $\rho_i(t)$ [N/m] is calculated as:

$$\rho_i(t) = 1/d_{h,i}(t) \tag{1}$$

where $d_{h,i}(t) = x_{i-1}(t) - x_i(t)$ which means that the headway $d_{h,i}(t)$ for a pedestrian $i$ at time $t$ is defined as the distance between the centers of pedestrian $i$ and his/her predecessor $i$-1. In addition, the index '$h$' represents that the distance is calculated based on the headway. Note that, not all of the three systematic errors mentioned above happen at the same time and deviate to different direction. Thus, the error of the headway calculated from the trajectories is within 0.1 m.

The speed $v_i(t)$ [m/s] of pedestrian $i$ is calculated as:

$$v_i(t) = \frac{x_i(t + \Delta t/2) - x_i(t - \Delta t/2)}{\Delta t} \tag{2}$$

where the time interval to calculate the speed $\Delta t = 0.8$s is adopted according to the analysis [33].

At the macro level, the average density $\rho(t)$ [N/m] and speed $v(t)$ [m/s] in the measurement area are defined as follows:

$$\rho(t) = \frac{\sum_i^{n(t)} \frac{l_i(t)}{l_m}}{l_m} \tag{3}$$

$$v(t) = \frac{\sum_i^{n(t)} l_i(t) v_i(t)}{l_m} \tag{4}$$

where $l_i(t) = \frac{1}{2}(x_{i-1}(t) - x_i(t)) + \frac{1}{2}(x_i(t) - x_{i+1}(t)) = \frac{1}{2}(d_{h,i}(t) + d_{h,i+1}(t))$ which means that $l_i(t)$ represents the personal space (the distance from the center of $x_i$ and $x_{i-1}$ to the center of $x_i$ and $x_{i+1}$) of pedestrian $i$ overlap with the measurement area. And $l_m$ represents the length of the measurement area, $n(t)$ indicates the number of pedestrians in the measurement area at time $t$.

### 4.1.2 The fundamental diagrams of elders

Although the speed calculated by equation (2) showed slight difference due to the different boundary conditions and geometry of the paths, we observed no significant difference in the fundamental diagrams from the macro level of area 1, area 2 and area 3 except for different fluctuations. The effects of individual differences and less sample points for Elders-08 and Elders-20 make it necessary to analyze the macroscopic fundamental diagrams in a longer measurement area. We decided to choose half of the circuit length as the measurement area and it is better to include all of the types of boundary. As a result, the measurement area [4 m, 17 m] is selected for all of the six runs for the following analysis to minimize the fluctuation in speed and density. Moreover, we only consider the data in steady state eliminating the initial and final transient to remove the impact of the start and stop phases.

The headway and speed distribution of each run are statistically analyzed in FIG.5 and the insets show the average and standard deviation values of the headway and speed in each run. It seems that these two variables fit with Gaussian distribution in different runs with different average values and deviations except for the speed of the run Elders-56, where two peaks can be observed obviously around $v_{i,t} \approx 0$ m/s and $v_{i,t} \approx 0.2$ m/s which represent the stopping and going motion states. This proves the existence of stop-and-go waves, which will be analyzed in detail in section 4.3. From the insets in FIG.5, the headway and speed are declining with the increase of the number of participants. However, the magnitude of the change decrease especially when the number of pedestrians is greater than 39 (the global density $>1.52$ m$^{-1}$).

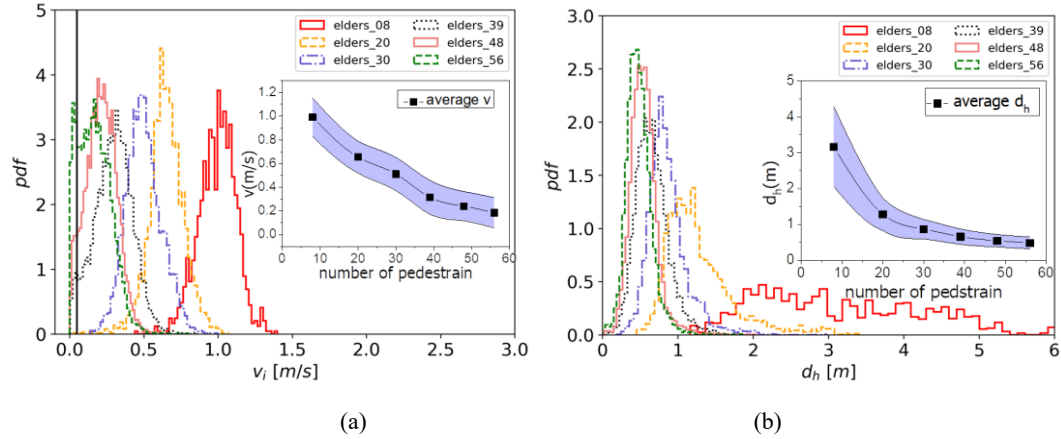

(a)  (b)

FIG.5. The distribution of the headway (a) and speed of the pedestrians (b) in different runs. The vertical line represents $v=0.05$ m/s. And the insets are the headway and speed versus the number of pedestrians.

FIG.6 shows the fundamental diagrams of single-file pedestrian flow for the elderly from the micro and macro methods respectively. The number of points in FIG.6 (a) are controlled to be consistent for different runs by changing the sampling frequency. While in FIG.6 (b) we selected the data from the relatively stable stage and only used one per 10 frames to guarantee the independence of the samples. Even if the maximal density is about 2.3 m$^{-1}$ at the macro level but reaches up to 5 m$^{-1}$ at the micro level, the similar trend can be observed from these two fundamental diagrams except for different resolutions, which is correspondence with the findings in [15] with different age compositions. It is worth noting that pedestrian movement properties of elders under higher densities are obtained in this study.

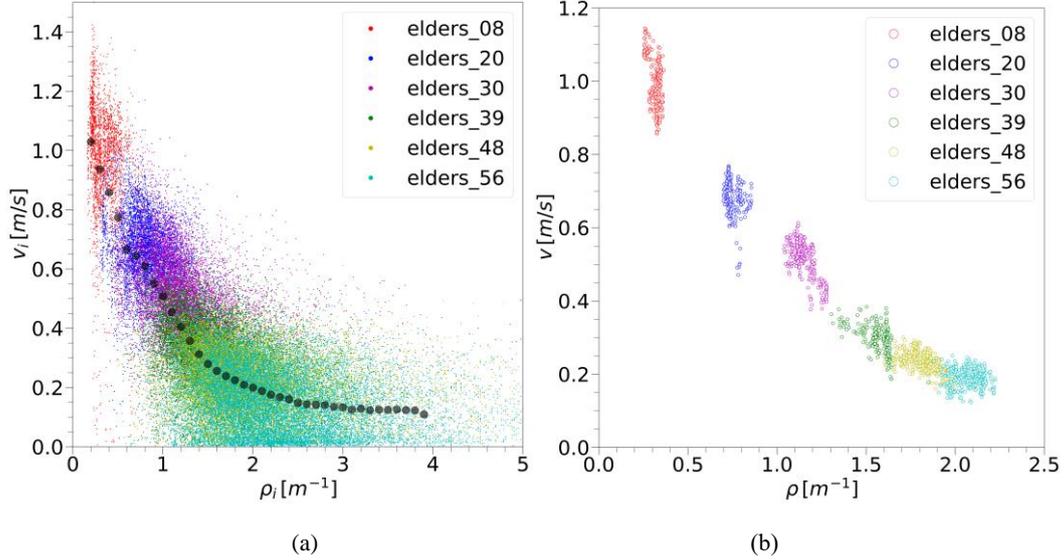

(a) (b)

FIG.6. The fundamental diagrams of the elders from (a) the micro level and (b) the macro level. At the micro-level, the density is obtained based on headway $d_{h,i}(t)$ while the speed $v_{i,t}$ is the instantaneous speed of each individuals as calculated in the equation (1) and (2). At the macro-level, the density and speed are calculated with the equation (3) and (4).

The scatter diagram FIG.7 (a) shows the relation between individual headway $d_{h,i}(t)$ and speed $v_i(t)$, while in FIG.7 (b) the fitting result is compared with the results in [14] with the same binning procedure as they mentioned. When the headway is greater than 4m, the speed of pedestrians fluctuate greatly and the data points are few, so we did not take these data into account in the fitting analysis. Three regimes (the strong constrained regime, weakly constrained regime and the free regime) can be observed for both the Chinese elders and the French students and they have same turning points ($d_h$=1.1 m and $d_h$=2.6 m). As interpreted in [14], pedestrian's movement is restricted when the headway is small in the constrained regime and they need to adjust the speed to avoid collision with the predecessor. While in the free regime, the headway is large enough and individuals can move with free speed. However, significant differences can be found quantitatively for these two groups. TABLE IV shows the slopes and the intercepts of the fitting lines. When fitting the lines, data from the vertical binning are considered in the strong constrained regime and data from the horizontal binning are considered in the other two regimes. Overall, the speeds of the elders are lower than that of the French students under the same headway especially in the free and weakly constrained regimes. When the headway is greater than 2.6 m, the speed is independent on the headway for both groups. The mean free speed is about 0.94 m/s for the elders and 1.15 m/s for French students respectively. The larger fluctuation for the elders in the free regime can be contributed to the large differences of their mobility and flexibility due to the larger age span. In the strong constrained regime, the main difference for the two group is that the slope for elders is bigger but the intercept is smaller. It means that the elders adapt their speed to the available space in front more strongly. In the weakly constrained regime, the speeds of pedestrians depend more weakly on the spatial headway than the strong constrained regime. It is interesting that the slopes of the two groups are similar which may mean that the influence of the headway on pedestrian speed is independent of age in this regime.

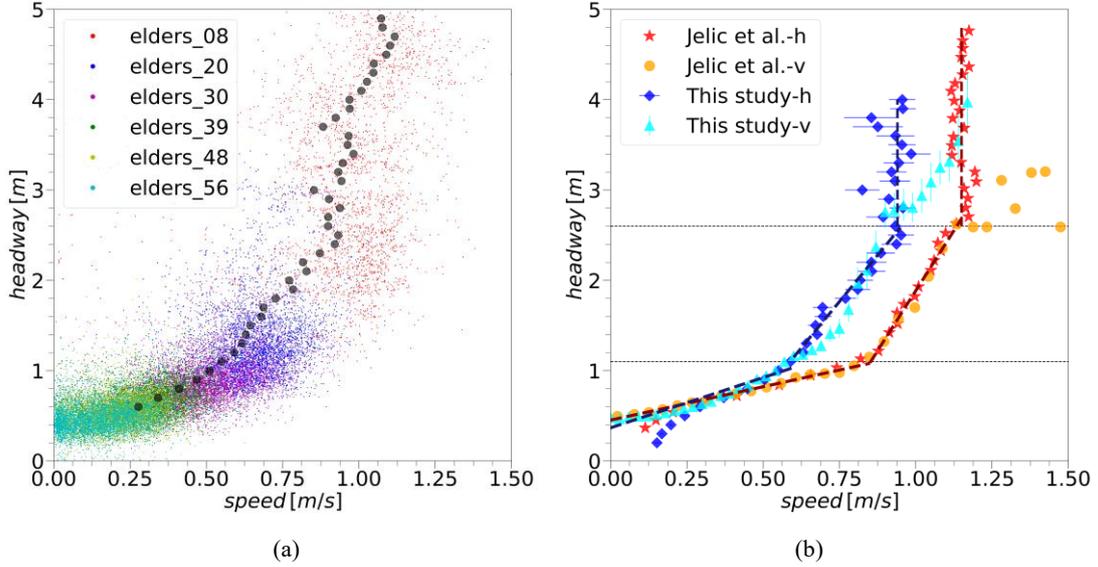

(a)          (b)

FIG.7. The relation between headway and speed. (a) The scatter diagrams of the instantaneous individual speed $v_i(t)$ and headway $d_{h,i}(t)$. (b) Comparison for the relation for the elders with that for the French students in [14] using the binning procedure, and the horizontal lines represent headway $d_h$=1.1 m and $d_h$=2.6 m.

TABLE IV. Slopes and intercept points between different regimes of speed binning.

| Regime | Strong constrained regime | | Weakly constrained regime | |
|---|---|---|---|---|
|  | Elders | French students | Elders | French students |
| Intercept (m) | 0.10 | 0.25 | - | - |
| Slope (s) | 1.11 | 0.74 | 4.41 | 5.32 |

**4.1.3 Comparison with different age compositions**

In this paper, five groups of pedestrians are compared to recognize the influence of age composition on the single-file movement. In order to make a clearer distinguish we listed the detailed information of each group in TABLE V.

TABLE V. The mean speeds with the standard deviation of different measurement areas in each run.

| Group | Elders | French students | Old adults | Mixed | Young |
|---|---|---|---|---|---|
| Source | Our experiment | Ref.[14] | Ref.[15] | Ref.[15] | Ref.[15] |
| Age | 69.7(52-81) | Not mentioned | 52(45-73) | - | 17(16-18) |
| Social status | Senior center | Students | Residential districts | Old adults+ Young | Students |

Firstly, we compare the relationship of speed and density (which is represented by the reciprocal of the headway) of elders group with that of the old adults and mixed group in [15] from the micro and macro aspects. As shown in FIG.8, the speed of elders is lower than that of old adults for the same headway, while it is difficult to observe the difference in the micro scatters of the elderly and the mixed group. For the convenience to compare the scattered points, we show the fitting results in Fig.8. It can be found that when the headway is longer than 0.5m, the mixed group have a higher speed than the elderly. The reasons can be explained by the mobility of these groups which can be reflected by the free speed and the familiarity between the pedestrians. The free speed of elderly pedestrians is 0.94±0.15 m/s calculated when the headway is between 2.5 m and 4.0 m and that of the old adults group (0.95±0.16 m/s) which is lower than that of the mixed group (1.05±0.16 m/s) in [15] which can be found in FIG.9. On the other hand, the similarity of the fundamental diagram for the elders and mixed groups may be due to not only the mobility but also the familiarity between neighboring pedestrians in the crowd. Compared to the old adults group and the young, the elders show obviously lower mobility but are familiar with each other. Whereas for the mixed group formed by the old adults and the young students, the age shows two peaks distribution (16 to 18 for the young and 36 to 75 years for the aged old adults with an average age of 52). The obvious different age span lead to different life styles and social status. With the mixed mode of the group, the students and old adults were arranged in the corridor alternately with a ratio of 1:1 in the mixed group while the elders distributed freely without strict division of their ages. The young people usually show respect and humility to the elders. They try to keep a long distance from the old people to avoid bumping into them and causing injury. As a result, the decrease of speed became very obviously and the different movement speed caused by the physical mobility becomes weak. Reflected on the fundamental diagram, the similarity is observed.

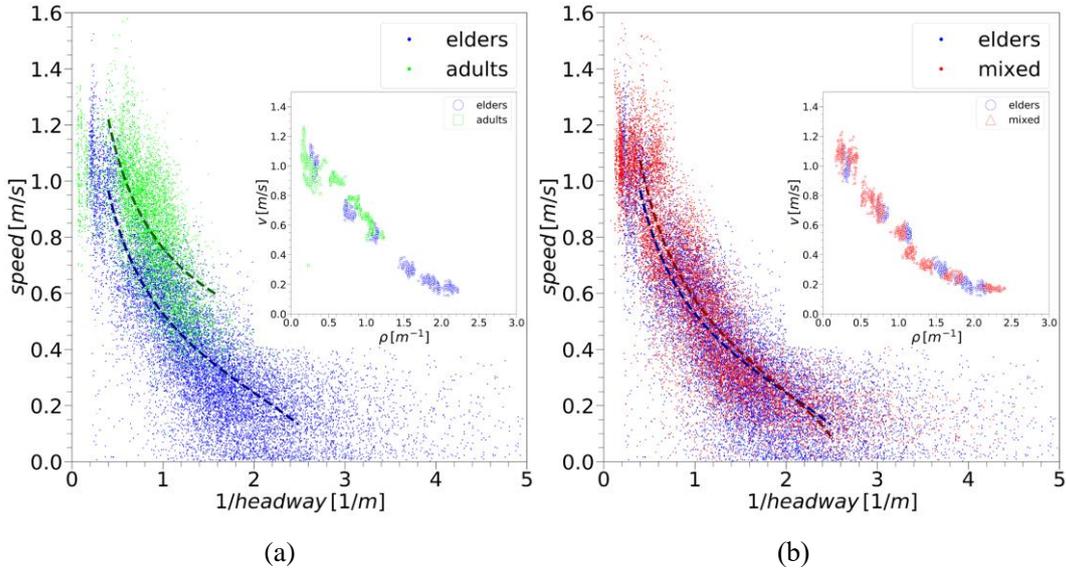

(a)                                           (b)

FIG.8. Compare the headway-speed of elders with the old adults group and the mixed group. The dashed lines are the fitting results of the scatter from the micro aspect. The fitting formula is $y = \sqrt{a/x+b}$ according to the nonlinear fitting results in Fig.9 (b).

To analyze the differences of headway and speed between different groups, we adopted the method of averaging the speed with distance headway 0.1 m as a segment to fit the three different age groups and then the method of linear fitting mentioned before is adopted. What should be mentioned is that the vertical binning is more suitable for the leftmost data (Fig.7). While for the convenience of comparison, we only exhibit the horizontal binning results in Fig.9. To exclude the effect of the leftmost data with large fluctuation, only data of the speed beyond 0.2 m/s are considered in the linear fitting. The results can be seen in FIG.9 (a) and we compare the linear fitting results of different groups in TABLE VI in which the results of mixed and young group are from [15]. As mentioned in [14], the slope has the dimension of time and represents the sensitivity to the spatial headway. It is also called as the adaptation time which is available to react before being at the minimal distance from the current predecessor. It can be seen that elders have the biggest slope and they need longer adaptation time than the other groups.

It is worth noting that the relationships of headway and speed for all of the groups including elderly group in our experiment, French students group in [14] and the mixed group in [15] have the same transition points ($d_h$=1.1 m and $d_h$=2.6 m). From the relationship of step length and headway in [17,34], we can see that the pedestrian stride is strongly influenced by the predecessors when the headway is smaller than around 1.1m, whereas the step length remain steady for $d_h > 1.1$ m. The transition around 2.6 m can be explained as following. Although the pedestrian stride is independent on the space in front for $1.1 < d_h < 2.6$ m, it is still necessary to adjust their movement rhythms to keep similar pace with others and avoid collisions. When the headway is larger than 2.6 m, however, the pedestrians can move freely and not care the space and other pedestrians in the movement direction. The same transition point of different age groups indicates that the effect of headway on pedestrians' stride law is independent of age.

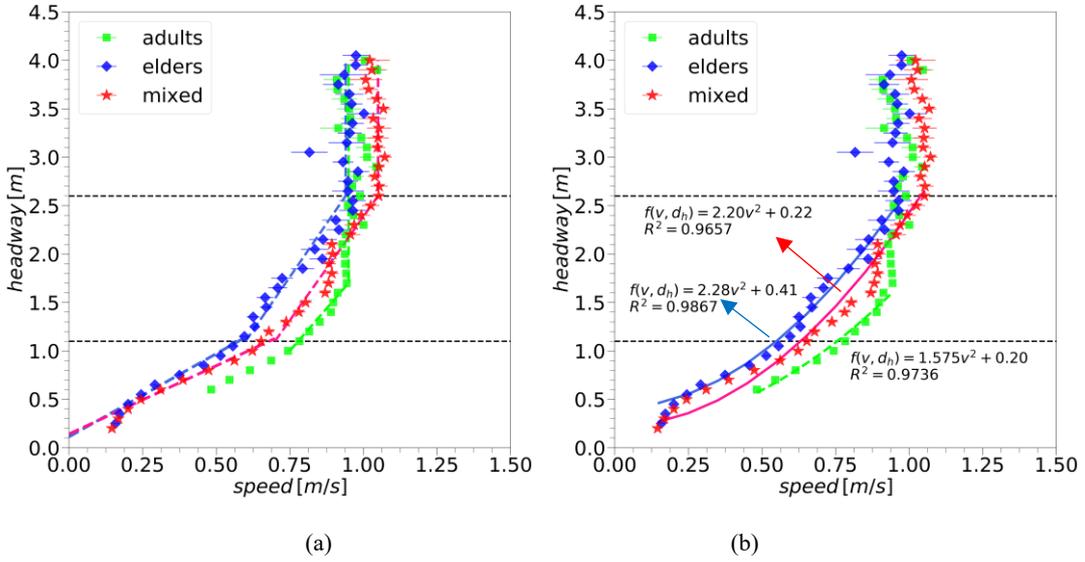

FIG.9 Relationship of the headway-speed using the horizontal procedure with the linear fitting (a) and the nonlinear fitting with quadratic function (b).

TABLE VI. Details of the linear fitting of different groups with different age compositions.

| Regime | Strong constrained regime | | | Weakly constrained regime | | |
|---|---|---|---|---|---|---|
| | elders | mixed | young | elders | mixed | young |
| Intercept (m) | 0.10 | 0.25 | 0.25 | - | - | - |
| $t_a$ (s) | 1.71 | 1.40 | 0.69 | 4.41 | 4.25 | - |

The movement of pedestrians in the single file can be compared to the car-following behavior in the traffic flow. It can be found in [35] that the safe distance for braking is proportional to the square of speed. While for pedestrian movement, the headway is equivalent to the safe distance of cars to avoid collisions. According to this consideration, method of nonlinear fitting is adopted and the quadratic function is selected to describe the relationship of headway and speed except for the free regime. As mentioned before, only data of the speed beyond 0.2 m/s are considered in the nonlinear fitting. FIG.9 (b) shows the results and we can see that these data have a high degree of fitting for $R^2$ is beyond 0.96. In the nonlinear fitting, the sensitivity increases with the increase of speed continuously rather than a mutation in the linear relationship and the elders have a larger increase range than the others. The relationship of headway-speed and adaptation time can be calculated using the expression in TABLE VII.

TABLE VII. Details of the nonlinear fitting of different groups with different age compositions.

| Groups | Elders | Old adults | Mixed |
|---|---|---|---|
| $f(v, d_h)$ | $2.28v^2+0.41$ | $1.70v^2+0.06$ | $2.20v^2+0.22$ |
| $t_a$ | $4.56v$ | $3.40v$ | $4.40v$ |

**4.2 Time-space diagram**

As the density increases, the so called stop-and-go is observed from the experiment. Reflected in the speed distribution (FIG.5 (b)), two peaks which represent the stopping and going states of pedestrians appears especially in the scenario of Elders-56 whose $\rho_g$ =2.18 m$^{-1}$. The two states seem to be separated around $v_i(t)$=0.05 m/s (the vertical line in FIG.6 (b)). As a result, we suppose that the pedestrians with $v_i(t)$<0.05 m/s are under stop state and those with $v_i(t)$>0.05 m/s are in going state to visualize it in the time-space diagram. The stop-and-go wave formed by a consecutive sequence of one or more stopping pedestrians. FIG.10 (a, b, c and d) shows the stop-and-go waves of the elders group in four different runs. No stopping occurred in the scenarios for $\rho_g$ <1.17 m$^{-1}$ (the number of the participants N<30), whereas obvious stop-and-go waves are observed frequently during the experiment for the global density $\rho_g$ >1.52 m$^{-1}$ (N =39, 48 and 56). Furthermore, the sizes and frequencies increase with the increment of the density. The speeds in which the stop-and-go wave propagates in the opposite direction to the movement of pedestrians are indicated by the absolute values of slope of the black lines. They are approximately 0.37 m/s for both runs of $\rho_g$ =1.87 m$^{-1}$ and $\rho_g$ =2.18 m$^{-1}$ (Elders-48 and Elders-56), while it is 0.5m/s for $\rho_g$ =1.52 m$^{-1}$ (Elders-39). Compare to that of the young and mixed group under high densities in [15], the stop state appears more frequently and lasts longer time at the same global density situations (see FIG.10 e, f, g and h). For further quantitatively comparison of the differences in stop-and-go among different age groups, we made statistics on the stop duration of stopping and going of the same number of people in different densities within 90s. The duration of stopping includes reaction time and waiting time. As shown in FIG.11, the stopping time of elders group is longer than the others obviously. This indicates to some extent that the density is not the only reason for the emergence of stop-and-go waves but the movement ability is an important factor.

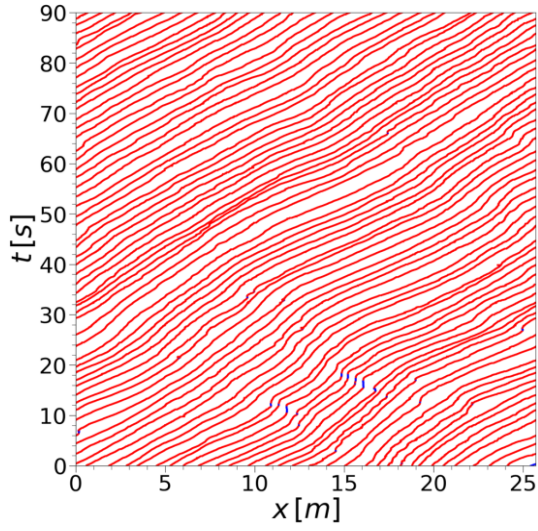
(a) Elders-30

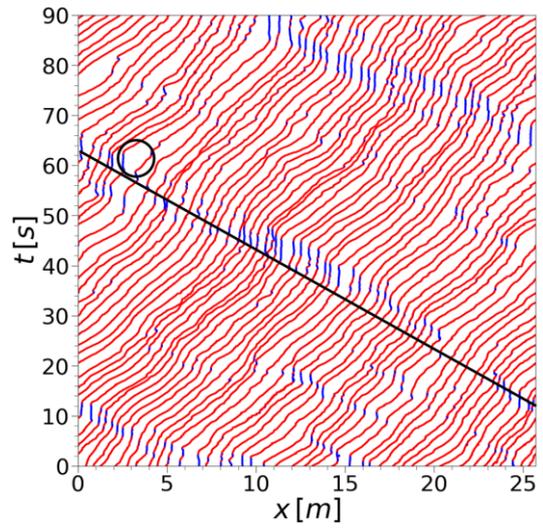
(b) Elders-39

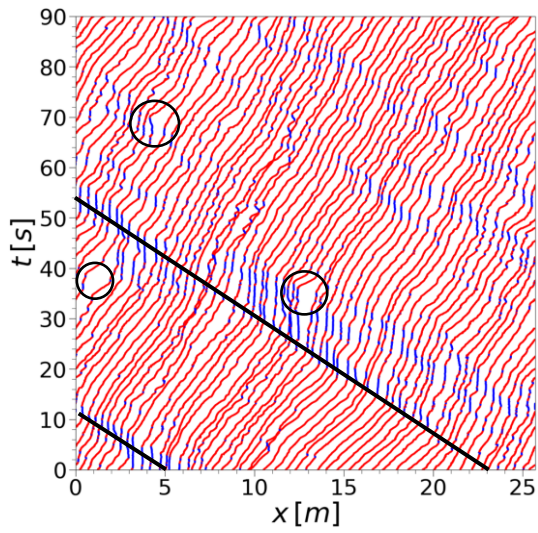
(c) Elders-48

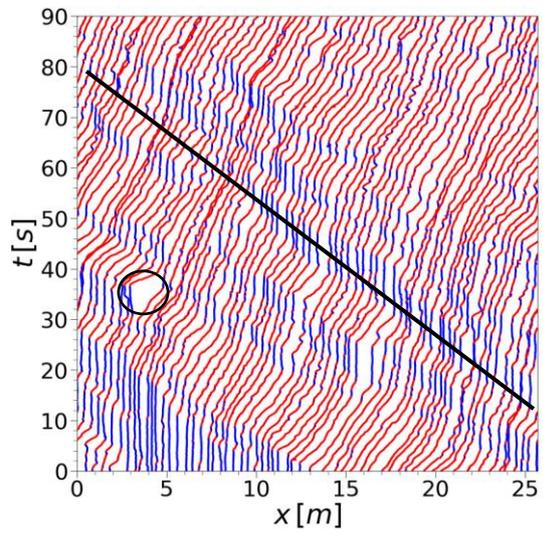
(d) Elders-56

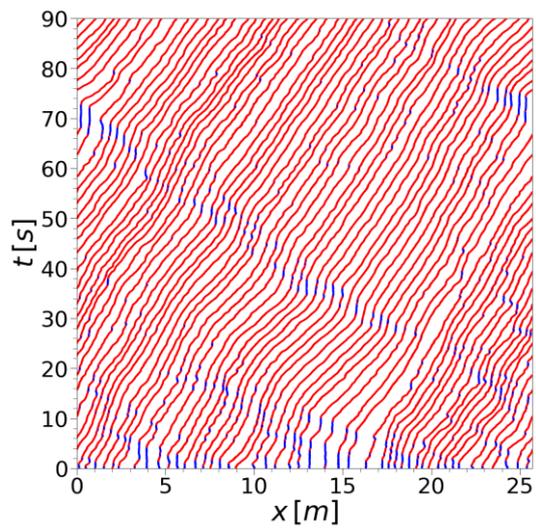
(e) Mixed-46

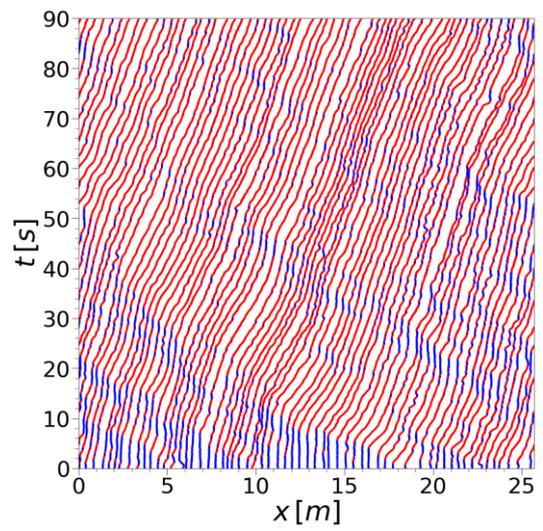
(f) Mixed-60

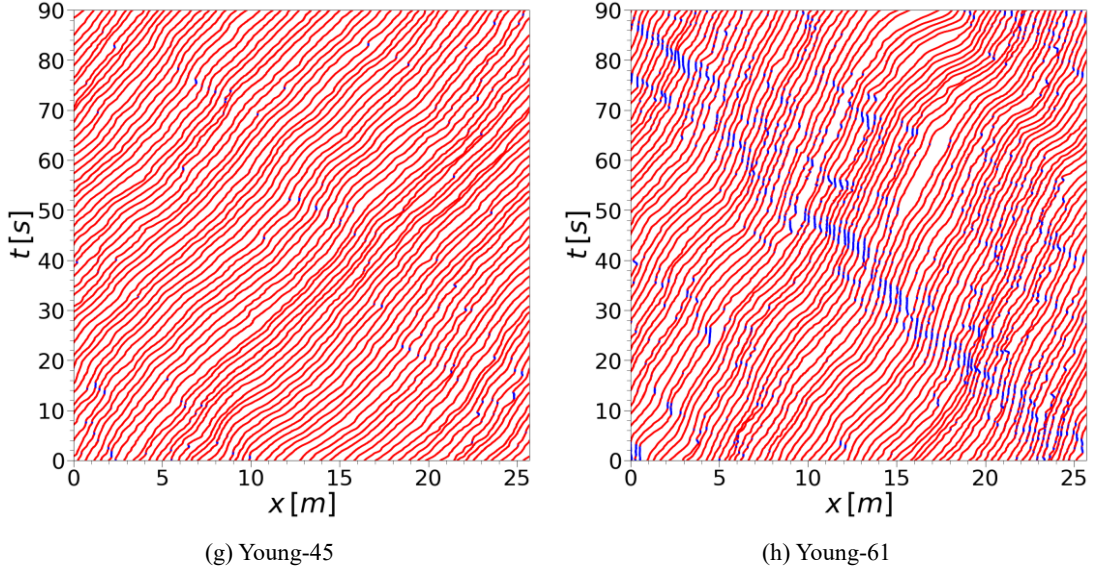

(g) Young-45  (h) Young-61

FIG.10. The time-space of the runs in our experiment and the previous experiment. The red indicates the speed higher than 0.05 m/s, while the small speeds are presented in blue.

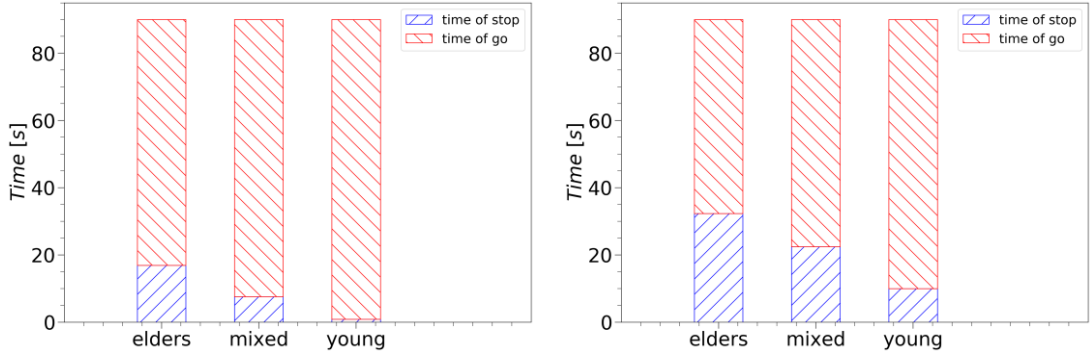

FIG.11. Quantitative statistics of the duration time of 'stopping' and 'going' states respectively of different age groups in 90s. The stopping time of elders group is longer than the others obviously. Left: 46 pedestrians are selected from these three groups of the run of elders-48, mixed-46 and young-50. Right: 56 pedestrians are selected from these three groups of the run of elders-56, mixed-60 and young-60.

Besides, if all pedestrians move in the single-file synchronously, the time-space diagram should be a group of parallel curves. However, some apparent gaps in the diagram of the elders group appears as marked by the black circles in FIG.10 (b, c, d). From the video recordings, it is observed that some elders did not start to 'go' synchronously with his or her predecessor after some stopping but waited several seconds till a certain amount of space in front and then start to 'go' again (e.g., the two marked pedestrians in FIG.12 (a)). We define this phenomenon as 'active cease' behavior of pedestrians. To study the 'active cease' behavior of pedestrians quantitatively, we calculate the headway of pedestrian in the moment of the transition from stopping to going under high density which is defined as $h_{s\text{-}g}$.

$$h_{s-g}(i, t_{s-g}) = x_{i+1}(t_{s-g}) - x_i(t_{s-g})$$

where $i$ is the selected reference pedestrian, $i+1$ is the predecessor of $i$, and $t_{s\text{-}g}$ is the changing moment of pedestrian motion state which needs to meet the following condition:

$$t_{s-g} = t\{v_i(t) < 0.05 \text{ m/s and } v_i(t+1) > 0.05 \text{ m/s}\}$$

where 0.05m/s is the critical speed for judging whether the pedestrian is stopping or going.

The statistical results of the $h_{s-g}$ of all of the pedestrian in runs of Elders-48 and Elders-56 within 90s are shown in FIG.12 (b) and TABLE VII displays the details. It can be found from the inserts that the $h_{s-g}$ is in accordance with the normal distribution with large fluctuation range (for the standard deviation is 0.17 m and 0.16 m). The $h_{s-g}$ of run Elders-56 (0.49±0.16 m) is shorter than that of Elders-48 (0.53±0.17 m) for p=0.000 in the T-test, where p is the probability that the difference between samples is due to the sampling error and p<0.05 represents a statistically significant difference. There were statistically significant differences. Pedestrians decrease the $h_{s-g}$ and follow their predecessors under the pressure from the rear pedestrian when the density increase, which is similar to the driving behavior. While the stopping duration time and the average number of $h_{s-g}$ per pedestrian increases as the density increases which can be observed from FIG.11 and TABLE VII respectively. This is also the reason why the higher the density is, the more obvious the stop-and-go wave will be. In addition, some outliers which are represented by the solid points beyond the upper edge are caused by that the headways of pedestrians with obvious 'active cease' behavior are significantly longer than that of others. The reason for this kind of 'active cease' can be explained from the aspect of the least effort principle. We know that the acceleration of a moving body is proportional to the force exerted on it in physics. Similarly, for pedestrians in the single file movement they should consume more energy to accelerate or decelerate during the alternation between stopping and going than in a steady state of motion like stopping for a long time or walking at a constant pace continuously. When pedestrians adopt "active cease" strategy, the frequency of transition from stopping to going is reduced for they reserve space for more continuous walking by increasing the stop time which leads to the reduction of energy consuming.

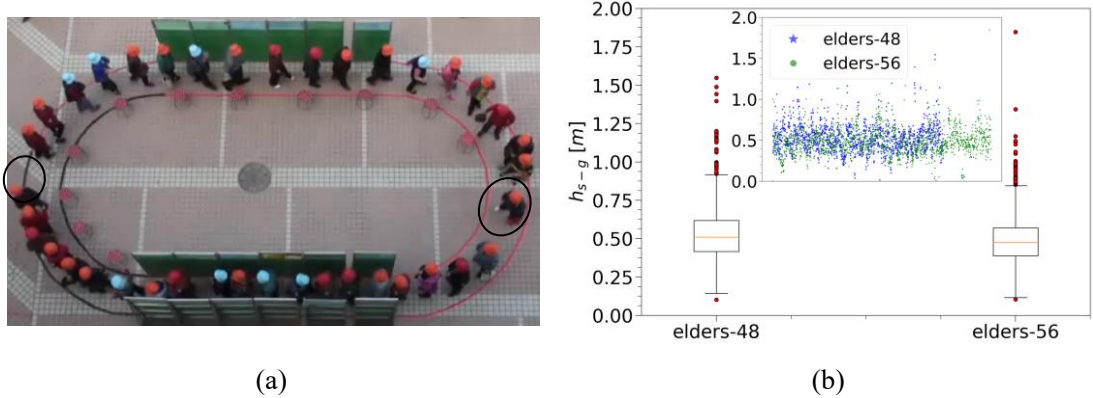

(a)          (b)

FIG.12. (a): These two pedestrians in the black circles are in the 'active cease' when the stopping is not due to the space limitation. (b): The statistical results of $h_{s-g}$ for Elders-48 and Elders-56, where the box charts are used to show the numerical distribution and the solid points filled with red represent the headway of 'active cease' pedestrian. In addition, the scatter and histogram inserts exhibit the raw data and its distribution.

To investigate the effect of age on the 'active cease' behavior, we analyze and compare the $h_{s-g}$ of elders, mixed and young groups respectively and the results can be found in FIG.13 and TABLE VIII. We can see that elders have the biggest mean and median value of $h_{s-g}$ compared with the other groups at the same density which indicates that the elderly need a longer headway at the moment of transformation from stopping to going. The longer $h_{s-g}$ make the elderly show up more obvious stop-and-go wave compared with the others. The per capita $h_{s-g}$ increases with density but the rate of increase is affected by age for it is 9.23%, 171.66%, 1034.75% for the elderly, the mixed and the young group respectively. Correspondingly, the per capita outliers of the mixed and young groups increased significantly, which makes the stop-and-go wave of these two groups more obvious at the higher density. In addition, the 'active cease' can be found in all of these three groups. The fluctuation of $h_{s-g}$ for the elderly group are larger than that of the other groups. What is interesting is that the percentage of bigger outliers decreases with the density increase for the elderly and mixed group. This also reflects the pressure effect from crowd density. What's more, the value of the highest outliers for the elderly is larger than that of the other groups. The reason may be that elders have weaker physical strength than the other which makes them more likely to reduce energy consuming by taking longer 'active cease' for longer continuous walking once start to move.

In conclusion, the stopping duration and times, and the number of the 'active cease' increase with the increasing density, while the value of $h_{s-g}$ and the proportion of the 'active cease' decreases. The magnitude of the change depends on age and groups with different age compositions show different tolerance for densities. It is speculated that the value of $h_{s-g}$ remains unchanged, the stopping duration increases but the number of the stopping times decreases when the density is greater than the critical. However, our experiment cannot reach higher density which requires further research.

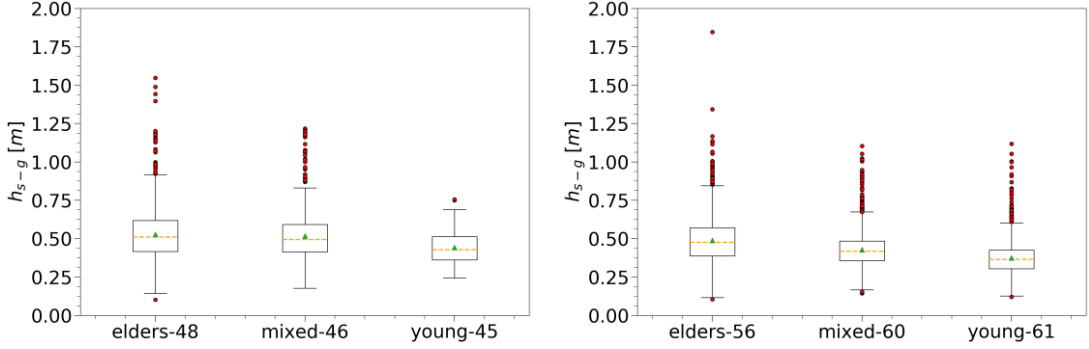

FIG.13. The statistical results of $h_{s-g}$ for groups with different age compositions.

TABLE VIII. Details of the $h_{s-g}$ of different runs with different age compositions.

| Runs | elders-48 | mixed-46 | young-45 | elders-56 | mixed-60 | young-61 |
| --- | --- | --- | --- | --- | --- | --- |
| Mean of $h_{s-g}$ (m) | 0.53±0.17 | 0.52±0.16 | 0.44±0.12 | 0.49±0.16 | 0.43±0.11 | 0.37±0.11 |
| Median of $h_{s-g}$ (m) | 0.51 | 0.50 | 0.43 | 0.47 | 0.42 | 0.36 |
| Number of $h_{s-g}$/ Per capita | 1362/28.38 | 982/21.35 | 127/2.82 | 1744/31 | 3438/58 | 1951/31.98 |
| Number of bigger outliers/ Per capita | 37/0.77 | 32/0.70 | | 42/0.75 | 87/1.45 | 63/1.03 |
| Percentage of bigger outliers | 2.72 | 3.26 | | 2.41 | 2.53 | 3.23 |

## 5. Summary

In this study, single-file walking experiment of elderly population in a 0.8 m wide circuit were performed under controlled conditions. The main aim of the study is to conduct comprehensive comparative studies among different age groups at high densities. 73 people over 60 years old participated in the experiment and the global density ranges from 0.31 m$^{-1}$ to 2.18 m$^{-1}$.

Closed boundary, semi-closed boundary and open boundary were set up to study their influence on pedestrian movement. Based on the trajectories extracted by the software *PeTrack*, it is observed that the pedestrians inclined to the side without wall in the semi-closed boundary area while it fluctuated greatly in the open boundary and concentrated more in the closed boundary. The speeds of these elderly pedestrians in the closed boundary were higher than that in the other parts under low density (Elders-08, Elders-20). The headway and speed distributions of the elderly pedestrians under different global densities are both normally distributed. Bimodal phenomena of the speed, which represents the existence of stop-and-go waves, was observed under high density.

By comparing the fundamental diagrams of elders in our experiment and that of the French young in [14], three linear regimes (free, weakly constrained and strongly constrained regime) are obtained in both studies. In the free regime (when the headway is larger than 2.6 m), Chinese elders have the mean free speed of 0.94 m/s which is lower than that of the French young whose free speed is 1.15 m/s. In the weakly constrained regime, the elders have a lower speed under the same headway. Interestingly, the small differences of the slopes indicates the neglectable influence of the headway on pedestrian speed. While in the strong constrained regime (when the headway is smaller than 1.1 m), the adaptation time of elders is longer than that of the young, which implies that elders adapt their speed to the available space more strongly. Then, we compared the fundamental diagrams of elders group with the groups in [15] from the micro and macro aspects. The elders showed a lower speed than the old adults under low density situations, while no obvious difference was observed between the elders and mixed group. The reason is explained from the heterogeneity of the crowd and the similarity among pedestrians. Besides, both linear and nonlinear fitting methods were adopted to analyze the relation between headway and speed.

From the time-space diagram, we observed the stop-and-go waves, which propagate in the opposite direction to the pedestrian movement with speeds roughly 0.37 m/s for $\rho_g \approx 1.87$ m$^{-1}$ and $\rho_g \approx 2.18$ m$^{-1}$. Compared to the results in [15], the stop-and-go waves occurs more frequently and lasts longer time in elders group, which indicates the density is not the only reason for the emergence of stop-and-go waves but the movement ability is an important factor. In addition, noticeable 'active cease' behavior were observed in the experiment. The duration and times of stopping, and the number of the 'active cease' increase with the increasing density, while the value of $h_{s-g}$ and the proportion of the 'active cease' decreases. The elderly take longer 'active cease' and need longer headway at the moment of motion state transformation ($h_{s-g}$), which is also the reason why the elderly show more obvious stop-and-go compared with the other groups under the same density.

The experiment in this paper supply abundant data for the movement dynamics of the elderly, which is useful for the establishment of computer models, the design and construction of pedestrian facilities that are friendlier to the elders in the future. Strictly divided age range of participants need to be considered to eliminate individual differences and to study the influence of the age on pedestrian movement characteristics in the future.

**Acknowledgement:**

The authors acknowledge the foundation support from the National Natural Science Foundation of China (Grant No. 71704168), from Anhui Provincial Natural Science Foundation (Grant No.1808 085MG217) and the Fundamental Research Funds for the Central Universities (Grant No. WK2320000040).